\documentclass[aps,prb,twocolumn,showpacs,floatfix]{revtex4}
\usepackage{graphicx}

\begin{document}

\newcommand{\zn}{ZnGa$_2$O$_4\,\,$}
\newcommand{\lc}{\lowercase}

\title{Effects of Fe substitution on the electronic, transport, and
     magnetic properties of ZnGa$_2$O$_4$: A systematic {\em ab-initio} study}

\author{Leonardo Pisani, Tulika Maitra, Roser Valent{\'\i}}

\affiliation{ 
Institut f{\"u}r Theoretische Physik, Universit{\"a}t Frankfurt, 
D-6 Frankfurt, Germany}

\pacs{pacs}
\date{\today}

\begin{abstract}

We present  a density functional study of  Fe doped into the tetrahedral
 and octahedral cation sites of  the wide band gap spinel ZnGa$_2$O$_4$.  
 We calculate  the electronic structure  for different substitutions  
  and discuss the magnetic and transport properties for each case considering
  different approximations for the exchange-correlation potential. We show that for certain
 doped cases, significant differences in the predicted behavior are obtained depending
on the exchange correlation potential adopted.
   Possible applications of the doped systems as magnetic semiconductors are outlined.

\footnotesize{}
 
\end{abstract}
\maketitle

\section{Introduction}

 In recent years,  spintronics  has emerged as one of the most studied
fields of research in the semiconductor physics because of the possibility
of exploiting both the carrier  spin and charge degrees of freedom 
 for the storage and transport of information in semiconductor devices\cite{wolf}.
Because of the already existing fabrication
technology for III-V semiconductors, the doping of these systems 
with magnetic ions (especially the cases of Ga(Mn)As and Ga(Mn)N)
is being intensively  investigated\cite{ohno}
 and a  large amount of
 theoretical and experimental work has been done in order to understand the
underlying mechanism for the ferromagnetism in these doped semiconductors.\cite{macdonald,pearton} 

 Recently, an alternative approach for this phenomenon
has been considered\cite{risbud} by  taking  a non-magnetic spinel semiconductor as host material
 and by doping it with  Fe.  The spinel structure (stochiometric formula AB$_2$O$_4$) with two
types of cation sites (tetrahedral A and octahedral B) offers new possibilities of obtaining ferromagnetic order by doping 
 one or both type of cation sites with magnetic ions. First experiments\cite{risbud} on the solid solution 
  [\zn]$_{1-x}$[Fe$_3$O$_4$]$_x$  with x=0.05,0.10 and 0.15 showed that long-range magnetic order
 is induced  with Curie temperatures up to 200K. Similar attempts by Krimmel {\it et al.}\cite{Krimmel_05}
on Fe$_{0.76}$In$_{2.17}$S$_4$ revealed, on the contrary, a spin glass state at low temperatures
 
 Along these lines, we investigate the electronic  structure of Fe doped \zn within density functional
 theory (DFT) with special emphasis on  the magnetic and transport properties.  Since the description
of these properties may be affected by the choice of the exchange correlation (XC) potential  considered
 within DFT, we perform a comparative study of the properties by considering two different XC 
approaches: the  Local Density Approximation (LDA)\cite{Perdew_92}  and the Generalized
Gradient Approximation (GGA) \cite{Perdew_96}. We observe that though the electronic structure
 of the parent compound \zn remains almost unaffected by the change of the XC potential chosen,
 the kind of Fe substitution (e.g. whether in A-site or in B-site) shows quite different magnetic and 
transport properties
 depending on the chosen XC functional.

We will show that the various doping options allow for a variety of interesting behaviors which
range from spin-polarized metal to the so-called 'transport half-metal'.  Though both
properties are  of
 potential interest for spintronic purposes, we will discuss that alternative choices of substituent
or host compound may improve the quality of these properties.

We have organized the paper in the following way.  In section II we present a study 
 of the volume optimization of the parent compound  within LDA and GGA, which sets the reference
 frame for the doped cases.     In the subsequent three sections (III, IV and V) we analyse  
 the electronic structure properties of Fe-doped \zn within both
 XC potentials in three different limits, namely,  Fe doped in the Zn position, Fe doped in the Ga position and
 Fe doped into both Zn and Ga positions respectively.
 Finally in section VI we present the summary of our calculations with a comparison
 to the available experimental data and
 conclude with suggestions for some possible future experiments.

\section{Parent compound}

\subsection{Computational details}

We have performed DFT calculations  using
 the full-potential linearized augmented plane-wave code WIEN2k \cite{wien2k}.
In our calculations, we chose APW+lo as the basis set.
The atomic sphere radii were chosen to be 1.8 a.u. for
Zn, Ga and Fe and 1.6 a.u. for O.  Expansion in spherical
harmonics for the radial wave functions were taken up to $l=10$ and the  charge densities and
potentials were represented by spherical harmonics up to $L=6$ .  For Brillouin-zone (BZ)
integrations  we considered a  60  ${\bf k}$ points mesh in the irreducible wedge  
and the modified tetrahedron method was applied.\cite{tetra}

\subsection{Volume optimization}

\begin{center}
\begin{figure}
\includegraphics[width=5cm,angle=-90]{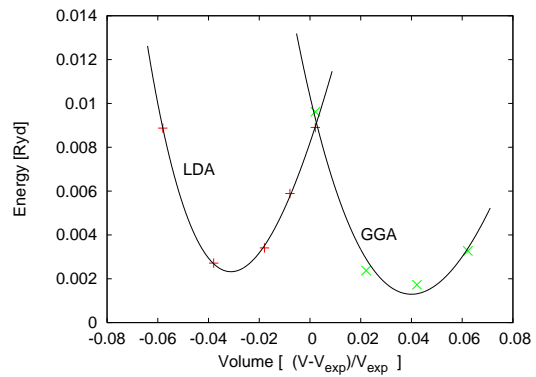}
\caption{(Color online) Volume optimization curves for LDA and GGA XC-functionals. The energy
scale has been made arbitrary for plotting reasons and the volume scale measures the relative
value respect to the experimental one.}
\label{volopt}
\end{figure}
\end{center}

ZnGa$_2$O$_4$ crystallizes in a normal spinel structure
(space group $Fd\bar{3}m$  (227)) with the primitive (rhombohedral) unit cell 
containing two formula units with Zn, Ga and O at the positions (1/8,1/8,1/8), 
(1/2,1/2,1/2) and (u,u,u) respectively (where u=1/4 corresponds to a perfect spinel structure).
Zn is sorrounded by a tetrahedral environment of oxygens while
Ga sits on an octahedral position.
In order to adopt a reliable approximation for the 
XC potential in the description of the parent compound, 
we performed a full volume optimization within
two widely used parametrizations of the XC functional: namely, 
the LDA\cite{Perdew_92} and  the GGA\cite{Perdew_96} approach. 
For each selected volume we also considered  the relaxation of the internal 
coordinate of oxygen (u parameter)  
as allowed by the symmetry in the space group 227,   by
using the damped Newton dynamics method,
until a force value smaller
than 1 mRyd/a.u. was reached.  The energy minimum of the
various selected volumes defines then the optimal volume.  Fig. \ref{volopt} shows the 
total electronic energy as a function of the volume of 
the rhombohedral  unit cell for LDA and GGA (V=${1 \over 4}$ a$^3$ where a is the
conventional cubic cell parameter).

The optimized lattice constant 
whose experimental value\cite{josties_95} is a=8.334 {\AA},
 is found to be underestimated in LDA by about 1$\%$
and  it is overestimated in GGA by about 1.5$\%$.
In Table \ref{dist} we show a comparison of tetrahedral (Zn-O) and octahedral
(Ga-O) distances between the optimized ones and the experimental ones.
Note that  the experimental values are underestimated by LDA and overestimated
by GGA. 
This kind of behaviour is generally not unexpected when adopting the 
LDA or the GGA approximations in the description of semiconductors. 
For instance,  in a recent {\em ab-initio} \cite{fang_02}
 study of the spinel ZnAl$_2$O$_4$,  the LDA lattice constant underestimation 
 was found to be of 1.1$\%$. The origin of the LDA and GGA shortcoming may be ascribed 
 to the inability of both approximations 
 to describe properly 
 the binding due to long-range forces like  the van der Waals interaction (apart from 
 the inaccuracy concerning exchange and correlation).
  In contrast, optimization of
 the   u parameter  for \zn  converges to  0.3861 within LDA and 0.3866 within GGA,
 both in  good agreement with the experimental value of 0.3867.

We have also estimated the residual and less important discrepancies due to
the neglect of the zero point  and thermal motion of the atoms.
Assuming the cubic spinel to be
an isotropic three dimensional harmonic oscillator with a spring
strength given by the curvature of the energy vs lattice constant function, 
we estimate the ground state length fluctuation to be 0.019 {\AA} within  LDA. 
This represents a deviation of about 0.2$\%$ with respect to
the classical adiabatic value.
The thermal expansion from zero to room temperature is expected to produce
an effect on the lattice parameter less than 0.1$\%$,
as   inferrable from the data of Josties {\em et al.} \cite{josties_95},
 therefore these corrections fall within the inaccuracy of the XC approximations used here.

From the fit of the energy-volume curves of Fig. \ref{volopt} to the
Murnaghan equation of state \cite{murn_44} we extract a bulk modulus for \zn
of 217 GPa within LDA and 146 GPa within GGA. Structure optimizations for the spinels 
ZnAl$_2$O$_4$  and ZnGa$_2$O$_4$ have been performed in the past
within the framework of the shell model \cite{pandey_99} providing
a bulk modulus of 273 and 237 GPa respectively. Recently Levy {\em et al.}\cite{levy_01}
have measured the bulk modulus of ZnAl$_2$O$_4$ to be 202 GPa well below the value
 obtained by the shell model (273 GPa).  Therefore, 
 assuming that the shell model bulk modulus
 is affected by the same amount of discrepancy for
  both  ZnAl$_2$O$_4$  and ZnGa$_2$O$_4$ and  since our calculated values for \zn  are lower
 than that of the shell model, we expect our results to be in a better agreement to
 experiment than the shell model result.

We  conclude from the previous comparison that the gradient correction to 
 LDA  as implemented in GGA doesn't produce a significant improvement
upon the LDA  itself as far as the host compound is concerned.  
However, in the following sections we will see that upon Fe substitution the LDA and GGA outcomes
may turn out to be quite different.


\begin{center}
\begin{table}
\caption{Optimized and experimental tetrahedral and octahedral distances.}
\vspace{.3cm}
\begin{tabular}{c c c c}
\hline
\hline
      \hspace{1cm}          &  LDA  & GGA  & exp.      \\
\hline
 Zn-O (${\AA}$)\hspace{.5cm}  &  1.943  & 1.997    & 1.974   \\
 Ga-O (${\AA}$)\hspace{.5cm}  &  1.975  & 2.018    & 1.991   \\
\hline
\hline
\end{tabular}
\label{dist}
\end{table}
\end{center}

\subsection{Electronic structure}

\begin{figure}
\includegraphics[width=7cm]{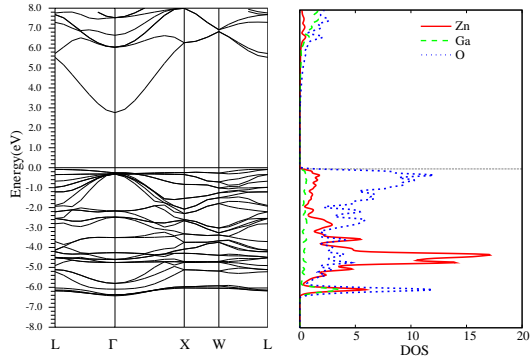}
\caption{ (Color online) Bandstructure and decomposed density of states for ZnGa$_2$O$_4$ within LDA. The path chosen
 in the Brillouin zone is  L=(1/2,1/2,1/2), $\Gamma$ = (0,0,0), X = (0,1,0) and W= (1/2,1,0) in units of 2$\pi$/a. The DOS is
 given in units of states per eV and per atom in all the figures.}
\label{parent}
\end{figure}

In this  section we present the density of states and bandstructure 
of the parent compound
within the  LDA and GGA approximations.
In Fig. \ref{parent} both properties are shown only for the LDA approach
since no significant modification is observed within GGA.  These results prove to be
 in good agreement with previous calculations\cite{sampath_99}.

The spinel ZnGa$_2$O$_4$ is a direct gap semiconductor with a measured band gap of about 4.0 eV
\cite{sampath_98}. At the 
$\Gamma$ point the band gap is calculated here to be 2.7 eV which reminds us of the renown tendency of LDA 
to underestimate the electronic band gap. No significant improvement on the gap value
is detected within GGA. 
The density of states (DOS) shows the valence band to be mainly of O character with a
peaked  Zn weight due to the full $d$ orbital shell. The Ga $d$ states are confined below -10 eV 
(not shown) and strongly atomic-like. 

It is important here to point out the dispersive feature of the
band structure at the $\Gamma$ point above the Fermi level which has mainly Ga and O  
character. The associated effective mass 
is calculated to be 0.28 times the electron rest mass and is rather isotropic; therefore 
doping with a magnetic ion may open the possibility for a high mobility electronic current which 
simultaneously might be fully spin polarized, as we will  discuss in the next paragraphs.
Finally, 
recalling that the DOS weights are given within the atomic sphere radii considered in the APW+lo basis,
the smallness of the weight of the dispersive $s$ band  at $\Gamma$  (Fig. \ref{parent} right panel)
is due 
to the large extension of the 4$s$ state outside the sphere (interstitial region).


\section{F\lc{e} doped into A position}

Substitution of Fe in one of the two Zn sites of the rhombohedral unit cell
removes the inversion symmetry from the space group $Fd\bar{3}m$ and reduces it
 to the maximal subgroup $F\bar{4}3m$ (space group 216). 
In addition, the original 8 equivalent oxygen positions are now split in two non-equivalent sets 
(Wyckoff position $16e$)\cite{ITA}. Defining  the doping concentration  
with respect to the total number of tetrahedral sites, 
this new structure corresponds to a 50$\%$  Fe doping.  
In a tetrahedral crystal field, the Fe $d$ levels are  split into
 energetically lower $e_g$ states and upper $t_{2g}$ states.

We have calculated the electronic structure for this system within the 
spin polarized versions of the LDA and GGA approximations. Both approaches lead to
 very similar results, therefore we  discuss here
 the LDA results and comment on quantitative differences with respect to GGA.

\begin{figure}
\includegraphics[width=7cm]{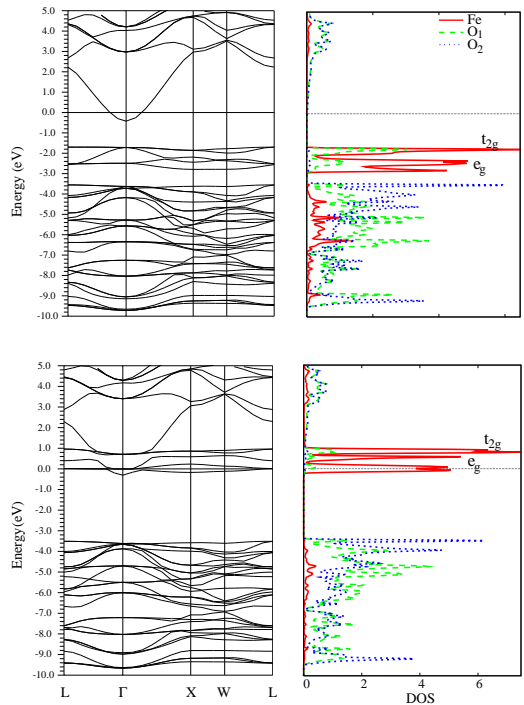}
\caption{(Color online) Bandstructure and density of states  for Fe doped into the A-site of ZnGa$_2$O$_4$ within
LSDA. Upper panels: majority states (spin up).  Lower panels: minority states (spin down).
Recall that the DOS weights are given within the atomic sphere radii.
Therefore the small $s$ weight at E$_F$ in the DOS panel is due 
to the large extension of the 4$s$ state outside the sphere (interstitial region).}
\label{FeA}
\end{figure}

In Fig.  \ref{FeA} the density of states
and bandstructure are shown for the spin-polarized LDA case (LSDA).
 Since no major modification with respect to the parent compound is observed
 as far as the Zn and Ga weights are concerned and
 the size of the band gap (2.7 eV) remains unchanged,
  we  show in the DOS  only the Fe and O weights  (Fig. \ref{FeA} right panels). 


Note the full inclusion of the iron antibonding $3d$ majority (spin up) states within 
the band gap of the parent compound (Fig. \ref{FeA} upper right panel),  
 while the $3d$ minority (spin down) states  hybridize with the highly dispersive  
 $s$ band at the $\Gamma$ point  (Fig. \ref{FeA} lower right panel). 
 In a previous work by Nonaka {\em et al.} \cite{nonaka} 
 the authors did a molecular orbital calculation of one Fe atom doped into the Zn position of a ZnGa$_2$O$_4$ 
 cluster and found the Fe $d$-states to be located in the band gap of the host cluster. 
 
Due to the absence of any $d$ character at the Fermi level for the majority states,
we may therefore argue that  the conduction properties of this compound
are strongly spin dependent. In fact,
the $s$ conduction in  the spin down channel may be  hindered by the presence
of the relatively localized $d$ states which are instead absent in the spin up channel.
The situation partially resembles the case of transition metal ferromagnets
like Iron, Cobalt and Nickel, where the Fermi level lies 
in the d$\downarrow$ band but the effective spin polarization in transport is positive, 
due to the more mobile  4s($\uparrow$) electrons.

The exchange splitting is comparable in size with the band gap (2.7eV)
and much larger than the crystal field splitting (0.8 eV).
The oxygen  hybridization is more pronounced in 
 the majority spin bands
than in  the minority ones, as  indicated by the larger weight of O
 in the majority states valence band. This is ascribed to the  $t_{2g}$
symmetry of the filled majority states which in a
tetrahedral ligand field have a $\sigma$-bonding  with oxygen, 
in contrast to the minority $t_{2g}$ states which are here almost empty.

According to an ionic picture, iron in A position is expected to assume
a 3d$^6$  (Fe$^{2+}$) configuration in a high spin state  with a  magnetic moment
of 4 $\mu_B$.
The calculated total magnetic moment per unit cell
is 4.01 $\mu_B$ and is distributed mainly among Fe, O1  and interstitial
contributions in a proportion 3.19, 0.08 and 0.47 $\mu_B$, respectively.
Recalling that the iron muffin tin radius considered in the calculation
 is 0.95 {\AA},   we expect the
interstitial magnetic moment to be mainly due to the $d$ orbital tails
leaking out of the atomic sphere. 

Calculations within GGA give a
 total magnetic moment per unit cell of 4.03 $\mu_B$,
and the  Fe, O1  and interstitial moments are 3.28, 0.07 and 0.46 $\mu_B$, respectively.
Looking at the Fe moment 
one clearly sees a larger degree of localization within GGA with respect to LDA.
To confirm this, we examine the total amount of charge in the atomic spheres  and we 
find that it is increased in GGA respect to LDA.
No other important differences have been detected.

Because of the highly localized nature of iron $d$ states,
one may argue about the importance of inclusion of
orbital on-site correlations on top of  a mere LSDA approach.
We  have performed a LDA+$U$ calculation 
and considered two different versions of the
double counting correction, namely AMF ("around mean
field") by Cyzyk and Sawatzky \cite{czyzyk_94} and FLL (full localized limit)
by Anisimov {\em et al.} \cite{anisimov_93} ($U$=4.5 eV and $J$=1 eV).
No relevant modification respect to the LSDA case has been noted in both cases.
In fact,  the LDA+$U$ method aims at producing  a fully orbitally polarized ground state
by opening a gap between the occupied and non-occupied states. 
In the present case both versions result only in increasing the crystal field 
splitting e$_g$-t$_{2g}$
of the minority states just above the Fermi level 
pushing the upper 3 t$_{2g}$ bands further above.

It is also
interesting to note that the ionic radii of Zn$^{2+}$ and Fe$^{2+}$
in a tetrahedral coordination are very similar, namely 0.6 and 0.63 {\AA}.
As a consequence, the global (lattice parameter) 
and local (tetrahedrally coordinating oxygen)
readjustment of the structure due to the substitution is expected not to bring major modification
to the electronic properties above discussed. This is in fact confirmed within  
 both  the GGA and LDA calculations where the force values acting on the O and Ga atoms
 are around 1 mRyd/a.u. and therefore negligible.
 This also confirms the minor importance of the differences between the two approaches
 when Fe is substituted in a tetrahedral site.
 In the next section we shall see that important differences appear when  
 Fe is substituted in an octahedral site.

To determine the ground state magnetic order of this systems we performed an antiferromagnetic calculation
doubling the present unit cell. Comparing the total energies of antiferro- and ferromagnetic configurations
we found that the ferromagnetic state is lower in energy by about 61 meV. This state was also predicted to be
the ground state in this doping limit from a proposed phenomenological model in an earlier communication
by two of us\cite{tm-rv}.

\section{F\lc{e} into B position}

Substituting iron into 1 of the 4 Ga sites of the primitive unit cell
of ZnGa$_2$O$_4$ causes the lattice symmetry to lower from  cubic (space group 227)
 to the rhombohedral space group  $R\bar{3}m$ (166). In this group, the 8 equivalent
oxygen positions of the parent group (Wyckoff position 32e)  
split into  2c and 6h positions, with the latter being
the oxygen atoms octahedrally coordinating iron. In an octahedral
 environment,  the $3d$ orbitals of Fe are split into energetically lower $t_{2g}$
and higher $e_g$ manifolds.

For the present case we find a significant difference between the LDA and GGA outcomes.
In fact we performed  volume optimizations for both approaches and we
find that LDA produces as a ground state 
 an intermediate spin state for Fe, i.e. the unit cell magnetic 
moment  is calculated to be 1 $\mu_B$. Therefore,  within a simplifying 
ionic picture,  only the  $t_{2g}$ manifold is populated with 3 electrons
in the up sector and 2 in the down one, resulting in a Fe$^{3+}$ oxidation state in the
low-spin state 1/2.

On the contrary,  GGA describes Fe in a high-spin state
with a unit cell magnetic moment equal to 5 $\mu_B$.
The reason for this strong difference may be connected to  the element Fe itself.
According to LDA, Fe is predicted  to be non-magnetic and with a fcc structure
while within GGA the correct magnetic bcc structure is found \cite{bagno_89}.
Therefore we may expect GGA to be more reliable than LDA for the B site doping.

Another reason in favour of GGA is the end compound zinc ferrite, 
ZnFe$_2$O$_4$,  where Fe replaces all the Ga atoms.
Recent bandstructure calculations \cite{singh} have shown that
the zinc ferrite behaves as a metal within LDA and as an antiferromagnetic insulator
with a small gap  within GGA. 
The transport gap has been measured to be  0.2 eV\cite{shabasy_97}
in support of the GGA outcome.

We can understand the strong discrepancy between LDA and GGA
 with the help of the Table 
\ref{dist} and recalling the typical Fe-O distances for the high-spin state 
of Fe (2.2-2.3 ${\AA}$) and for the low-spin state  (1.8-1.9 ${\AA}$).
We see that the Ga-O LDA distance is a low-spin distance, therefore
after substitution and after volume optimization 
the Fe-O distance, which increases only by 1-2$\%$,
will remain of the low-spin state type.
On the contrary the GGA Ga-O distance is much closer to the high-spin region
and thus after volume optimization Fe will be in a high-spin state.

\begin{figure}
\includegraphics[width=7cm]{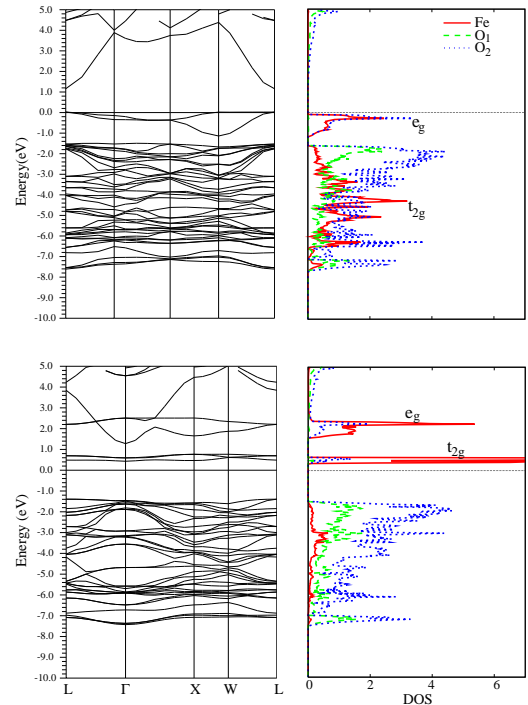}
\caption{(Color online) Bandstructure and density of states  for Fe in the B-site of ZnGa$_2$O$_4$ within
GGA. Upper panels: majority (spin up) manifolds.  Lower panels: minority (spin down) manifolds. The path
chosen in the Brillouin zone is $\Gamma$ = (0,0,0), L = (0,1/2,0),  Z= (1/2,1/2,1/2) and F=(1/2,1/2,0) in units of the trigonal reciprocal lattice vectors.}    
\label{FeB}
\end{figure}

In Fig. \ref{FeB} we present
the GGA  bandstructure and DOS. 
As in the previous case, the 
band gap of the parent compound remains approximately unchanged 
($\sim$ 2.7 eV). The majority spin $d$ states of iron (Fig. \ref{FeB} upper panel)
are fully occupied, 
of which the t$_{2g}$ subband completely hybridizes with the oxygen valence band
and the e$_g$ antibonding states set the Fermi level.
Concerning the minority states (Fig. \ref{FeB} lower panel), those antibonding with oxygen are completely empty implying that 
the Fe is in an oxidation state $3+$. 

From Fig.  \ref{FeB}  we find  thus a zero-temperature insulator with a band gap of 0.4 eV.
It is interesting to compare this band gap with that of the end compound ZnFe$_2$O$_4$  where within GGA  it 
is calculated to be one order of magnitude less\cite{singh}.
This can be ascribed to the direct cation-cation  bonding in the case of ZnFe$_2$O$_4$.
Since the Fe atoms are at the center of edge sharing oxygen octahedra,
 the t$_{2g}$ orbitals are able to directly overlap among each other
producing a widening of the t$_{2g}$ bands due to hybridization. 
This causes the minority
 t$_{2g}$  bands just above the Fermi level to become closer to E$_F$ 
 with a consequent reduction of the band gap. 
The spin exchange splitting for the present doped system is of about $\sim$2.5 eV, similar to the previous case,
but the crystal field splitting is around $\sim$1.8 eV, 
more than the double as for the previous case.

 The unit cell magnetic moment which is calculated to be 5.0 $\mu_B$ within GGA,
is distributed mostly among the Fe muffin tin sphere, interstitial region
and oxygen muffin tin sphere in a proportion  3.93 , 0.40  and 
0.10 $\mu_B$ respectively.

Note that in this case,
 at  the $\Gamma$ point the $d$ band in the minority sector 
remains above the Fermi level
due to  the oxidation state of Fe (3+) which leaves the minority
 antibonding $d$ states unpopulated. Moreover, due to the larger crystal field effect, 
 only the e$_g$ bands hybridize with the 4s band at $\Gamma$ while the t$_{2g}$
states  are pushed down just above the Fermi level. Therefore, if we were to reproduce
a situation similar to the A site case (see previous section)
 in which the $4s$ band should be at the Fermi level, 
we would need a transition metal in a $3+$ oxidation state with 9 electrons in the $d$ shell,
namely Zn$^{3+}$, which doesn't exist.


In order to determine the ground state magnetic order,  we compared
the total energies of ferromagnetic and antiferromagnetic alignments
of the iron spin.
 For this purpose
   we doubled the primitive
    unit cell into a tetragonal one containing 4 Zn, 6 Ga,  2 Fe and 16 O atoms.
       We have then
       considered an antiferromagnetic arrangement, where 
       the spins of the two non-equivalent Fe atoms in the tetragonal unit cell are 
       antiparallel to each other. This type of magnetic
       arrangement gives us planes of ferromagnetically aligned ions with an interplane 
       antiferromagnetic coupling.
       Comparing the total energy between this alignment and the ferromagnetic one,
       we find the former already to be lower in energy by 1 eV with respect to the latter.

In conclusion, we find that doping Fe into Ga site in a concentration of 25$\%$ results in an
 antiferromagnetic semiconductor at zero temperature which could show conducting properties with
  increasing temperature. 
 Namely, the t$_{2g}$ minority states can be thermally populated and produce a spin-polarized current. But, 
 due to flatness of the t$_{2g}$ bands just above the Fermi level the mobility of the carriers 
 would be very low. 
 On the contrary, due to the strong difference 
 in the hybridization of $e_g$ and t$_{2g}$ states with oxygen,  the $e_g$ bandwidth is 
 much larger than the t$_{2g}$ one, as clearly seen from the bandstructure shown in Fig. 
 \ref{FeB}. 
 Therefore,  setting the Fermi level into the majority $e_g$ manifold  
would induce  a  half-metallic behaviour  with a fully spin-polarized current. This can be
achieved by doping for instance with Mn instead of Fe.

\section{F\lc{e} doped  into A and B positions}

\begin{center}
\begin{figure}
\includegraphics[width=9cm,angle=0]{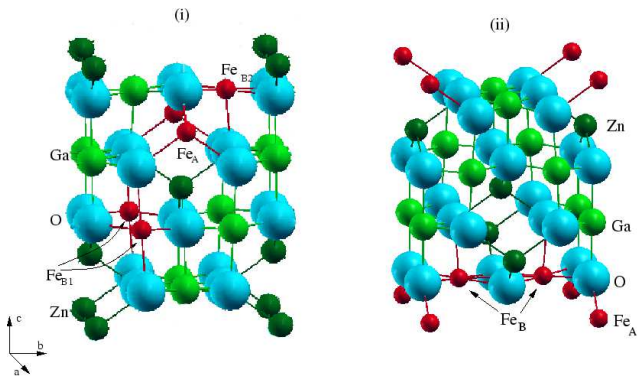}
\caption{(Color online) Crystal structures considered for Fe doped in A and B position
 in a concentration of 25 \% (see text).  In both cases Fe$_A$ 
is situated at the corners of the unit cell and in case (i)  
the Fe$_B$ atoms (Fe$_{B1}$-Fe$_{B1}$/Fe$_{B1}$-Fe$_{B2}$) distance is 5.83 ${\AA}$ while in case (ii)
the Fe$_B$ atoms distance is 2.92 ${\AA}$. }
\label{struct}
\end{figure}
\end{center}

The third case that we considered in this study is  the
doping of iron in both sites A and B in a ratio 1:2 (Fe$_A$:Fe$_B$ ) and in a total concentration of
25 $\%$.
Accordingly, we double the rhombohedral primitive unit cell of the parent compound
into a tetragonal unit cell containing 4 Zn, 8 Ga and 16 O.
Defining the doping concentration with respect to the total number of 
occupied cation sites,
 25 $\%$  doping of Fe (in both A and B sites) corresponds
to substituting 1 Zn with 1 Fe and 2 Ga with 2 Fe in the tetragonal unit cell.
Two inequivalent structures turn out to be consistent with this sort of doping 
(Figs. \ref{struct} (i) and (ii)), both belonging to the space group P$m$ (6).
In the structure (i) the Fe$_{A}$  and Fe$_{B2}$ atoms form zigzag chains with  
O atoms in between them (e.g. Fe$_{A}$-O-Fe$_{B2}$-O-Fe$_{A}$) along the $a$ direction and the  Fe$_{B1}$ atoms 
are connected to the  Fe$_{A}$ through O along the $c$ direction.
  The angles Fe$_{A}$-O-Fe$_{B1}$ and  Fe$_{A}$-O-Fe$_{B2}$ are equal to 121$^\circ$ and there
is no direct path between Fe$_{B1}$ and Fe$_{B2}$.
 In the structure (ii) the two Fe$_B$ are crystallographically
 equivalent and their distance is much shorter (2.92 {\AA})  than in the case (i) (5.83 {\AA}),
 therefore  direct  metal-metal  bonding is realized in this situation.

 To clarify the role of the on-site correlation 
  typical of the transition metal ions  for the present case 
  we have performed calculations within the  LDA+U potential, 
  besides the GGA one.
 We will present here only the LDA+U\cite{anisimov_93} results 
 since no major differences have been detected from the GGA outcome.
 (The LDA calculation has not been taken in consideration because of
   the failure of LDA   in describing Fe in the B case.)
   The values considered for the Hubbard and exchange parameters 
   U and J are 4.5 and 1 eV, respectively.

\begin{center}
\begin{figure}
\includegraphics[width=7cm]{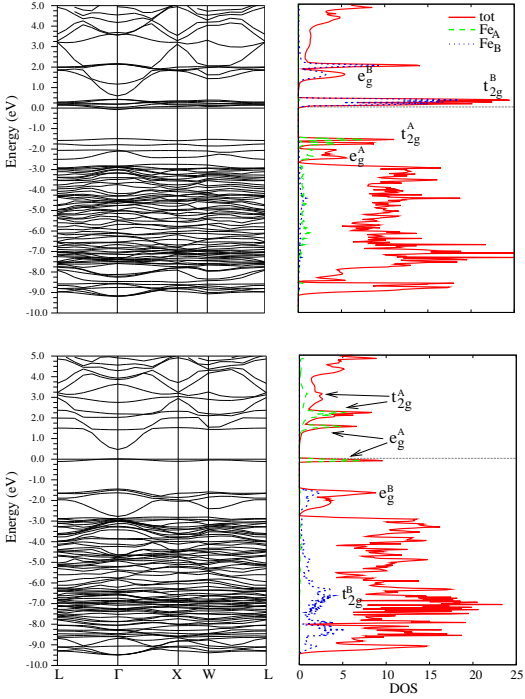}
\caption{(Color online) Bandstructure (left panel) and total and decomposed   density of states (right panel) 
(color online) for the
structure of case  Fig. \protect\ref{struct} (i) within LDA+U with $U=4.5$ eV
and $J=1$ eV. Shown are the results for
 the up spin manifold (upper panel of the figure) and the down spin manifold (lower panel of the figure).
 }
\label{dosa2b}
\end{figure}
\end{center}

In Fig. \ref{dosa2b} we present  the spin polarized density of states (DOS)
and related bandstructure within L(S)DA+U for the structure of Fig. \ref{struct} (i).
We have adopted a ferrimagnetic alignment of the Fe spins, specifically the
Fe$_A$ spins are  up and Fe$_B$ spins are down.  Therefore the majority states of Fe$_A$ are
spin up and the majority states of Fe$_B$ are spin down.   The upper
panel of the Fig.  \ref{dosa2b},  which shows the bandstructure and DOS of the spin up
species,  will contain the  Fe$_A$  majority states and the  Fe$_B$  minority states, while
the lower panel of  Fig.  \ref{dosa2b} (spin down species) will contain
the  Fe$_A$ minority states as well as the Fe$_B$  majority states. 

\begin{center}
\begin{figure}
\includegraphics[width=6cm,angle=0]{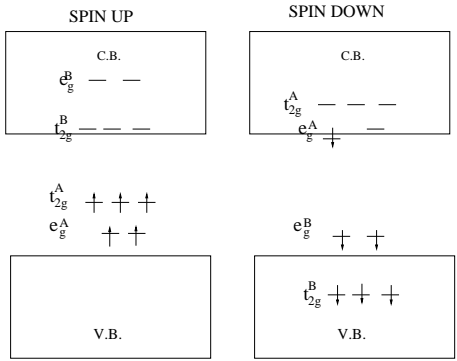}

\caption{Schematic picture showing the relative positions of Fe$_A$ and Fe$_B$ levels in the bandstructure
of the doped \zn for the crystal structure of Fig.  \protect\ref{struct}.  C.B. and V.B. denote
 the conduction and valence bands of the host compound.}
\label{ionic}
\end{figure}
\end{center}

In Fig. \ref{ionic} we present a simplified schematic energy level to show the approximate
location of the $d$-bands of Fe$_A$ and Fe$_B$.

 As  the two Fe$_{B1}$ and  Fe$_{B2}$  bands have rather similar features,
we show in Fig. \ref{dosa2b} (right panel)  the sum of the two Fe$_B$ weights.
In the energy range from -10 eV to about -3 eV the host compound valence band
is found to have  predominant oxygen character. In this  range 
the Fe-O hybridization  
occurs mainly in the   Fe$_B$ spin down bands (see lower panel of Fig. \ref{dosa2b})
 and less importantly in the Fe$_A$ states.
In the interval from -3 eV to -1 eV the impurity bands are clearly visible:
the spin up bands (see upper panel)  show the Fe$_A$ e$_g$ nonbonding and the t$_{2g}$ antibonding states
in the order of increasing energy,  and  the spin down bands (lower panel)
are mainly  Fe$_B$ e$_g$
antibonding states spanning the same energy range. At the Fermi level (see also Fig. \ref{ionic}) the 
 Fe$_A$ e$_g$ spin down  band (lower panel), due to the lowering 
of the  point symmetry at the A site, splits into an almost fully occupied band and an empty $e_g$
band 
at about 1.5 eV above the Fermi level.  In the upper panel,
 the Fe$_B$ t$_{2g}$ spin up bands 
are found slightly occupied and  above them we distinguish the empty Fe$_B$ e$_g$ states.
 In the same energy range  we find  in the lower panel the split 
 e$_g$ spin down bands and the 
t$_{2g}$ bands of Fe$_A$.  The effect of U will be discussed in the next paragraph.

According to an ionic picture, the oxidation states of Fe$_A$  and Fe$_B$
would be 2+ and 3+ respectively. However,  due to the small overlap
of  opposite spin bands at the Fermi level,  the ionic values are slightly
changed (increased for Fe$_A$ and reduced for Fe$_B$). 
In fact the magnetic moment of the unit cell is found to be -5.97 $\mu_B$,
while in the case of an ionic picture  it would be -6 $\mu_B$, 
since the magnetic moment of Fe$_A$  and Fe$_B$'s are respectively
4 $\mu_B$ and -5 $\mu_B$ (4+2(-5)).
A comparison with magnetite, which represents the end compound for Fe doping in both
A and B sites and it has a fully inverted spinel structure, i.e. (Fe$_A$$^{3+}$)(Fe$_B$$^{2+}$ Fe$_B$$^{3+})$O$_4$
tells us that in the compound (i) the Fe valence states are beginning to invert,
namely the  Fe$_A$ increases its oxidation state towards 3+ and Fe$_B$ decreases it to 2.5+.

An important point concerning the bandstructure shown in Fig. \ref{dosa2b} is
the role of the Hubbard U in the semimetal picture.
Since the Hubbard U represents the on-site Coulomb correlations
and no intersite Hubbard V has been included in our calculations,  no effect 
 on the electronic structure at the Fermi level of Fig. \ref{dosa2b} is expected
by varying U
since it contains two sets of bands originating from two different sites.
In fact, we performed a further calculation changing the  U value of Fe$_A$ and Fe$_B$
and, as expected, only a relative shift respect to the Fermi level 
is found for occupied (downwards shift) and unoccupied (upwards shift) states
leaving the relative position of the up and down bands at the Fermi level unchanged.
The main effect of the Hubbard U is thus to open a gap between occupied 
and unoccupied states and to  increase the crystal field splitting 
 which is already  taken into account  by an only LDA description. 
Therefore in the case of Fe$_B$ sites,  the Hubbard U just adds to the exchange 
interaction constant since the down states are fully occupied and the up states
are almost completely empty, while  in the case of Fe$_A$  site due to
the presence of Fe$_B$ as second nearest neighbour, which further splits the degeneracy 
of the e$_g$ bands,   the Hubbard U transforms  this split into a sizeable gap.
In a pure LSDA treatment the Fe$_A$ e$_g$ down  bands wouldn't show a relevant
splitting, thus overlapping with Fe$_B$ t$_{2g}$ up bands and the conduction band
wouldn't have an almost pure spin polarization nature as in the present case.

Another important effect to take into account is 
 the unit cell expansion or contraction due to the replacement of some of the 
Zn and Ga atoms with Fe. 
In this case we cannot  describe Fe$_A$ and Fe$_B$ as in a 2+ and 3+
oxidation states, respectively, since the compound is partially inverted,
and thus we cannot  expect minor changes of the structure
due to the similar ionic radii of  Zn$^{2+}$ with Fe$^{2+}$
and Ga$^{3+}$ with Fe$^{3+}$, as seen in the previous cases.
To have an idea of the structural modification, 
we recall that the ionic radius of Fe$^{3+}$ in A site is
0.49 {\AA}  and it is 0.78  {\AA} for Fe$^{2+}$ in B site,
while the Zn$^{2+}$ and Ga$^{3+}$ ionic radii are respectively
0.6 {\AA} and  0.62 {\AA}.
We shall discuss the atomic forces for the present structure at the end of this section.

 Finally, we can attempt a rough estimation of the ferrimagnetic ordering temperature
 by exploiting the analogy with magnetite Fe$_3$O$_4$.
The latter is known to have the highest ferrimagnetic temperature of 858 K.
 Considering the expression k$_B$T$_N$$\simeq$4$\sqrt 2$J$_{AB}$S$_A$S$_B$ \cite{kouvel},
where the coupling constant J$_{BB}$ between two B sites and
between two A sites  (J$_{AA}$) has been neglected with respect to the magnitude of 
  J$_{AB}$, 
one obtains a good estimation of the N{\'e}el temperature for magnetite
when  J$_{AB} \simeq 19 K$ .
In our case we assume a simple multiplicative rescaling effect of 
the magnetic ion concentration on the geometric factor 4$\sqrt 2$
and that the exchange constant doesn't
depend on the absolute spin  value of the two Fe. 
Calculating the total energy differences between the ferromagnetic and ferrimagnetic
alignment of the Fe spins we obtain an exchange constant J$_{AB}$ value of 25 K
which by means of the above formula provides us with a transition temperature of 260 K.

\begin{center}
\begin{figure}
\includegraphics[width=7cm]{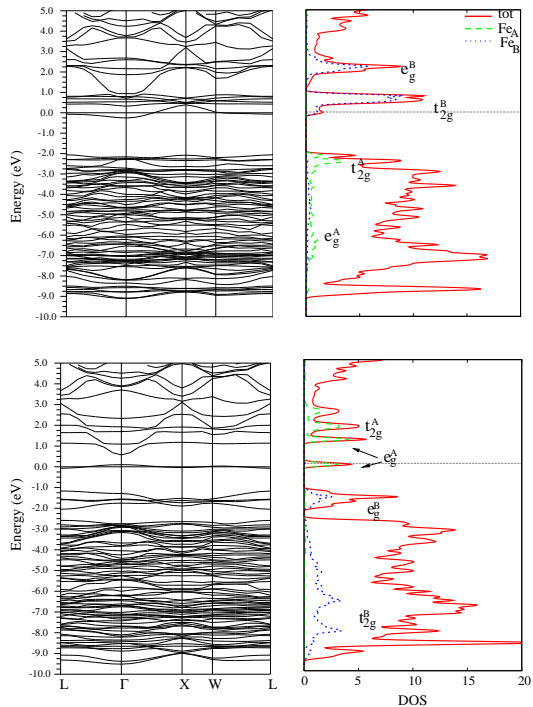}
\caption{(Color online) Bandstructure (left panel) and total and decomposed   density of states (right panel) for the
structure of case  Fig. \protect\ref{struct} (ii) within LDA+U  with $U=4.5$ eV
and $J=1$ eV. Line description as in Fig. \protect\ref{dosa2b}.}
\label{dosa2bii}
\end{figure}
\end{center}

We now discuss the case Fig. \ref{struct}(ii). 
It is important to note that the driving mechanism here is due to the  Fe$_B$-Fe$_B$ bonding
while  in case (i)  Fe$_B$ are connected to Fe$_A$ through oxygen
and  no direct  Fe$_B$-Fe$_B$ bonding is realized.

Fig. \ref{dosa2bii} shows the total and decomposed 
 densities of states within L(S)DA+U of both spin species.
 The magnetic ground state is found to have  antiferromagnetic
 order between Fe$_A$ (majority spin $\uparrow$ ) and  Fe$_B$ (majority spin $\downarrow$).
 The  spin up bands (upper panel) of Fe$_A$ and  Fe$_B$ are completely filled 
 and confined in the valence band  which has a charge gap of about 1eV 
 with respect to the dow spin manyfold.  The  spin down bands (lower panel) are at the Fermi 
 level for both Fe$_A$ and  Fe$_B$, indicating a
 mixed valence oxidation state. Similar to case (i), the point symmetry at
 the Fe$_A$ site is reduced by the presence of next nearest neighbor Fe$_B$ 
  which leads to the splitting of
 the  Fe$_A$ $e_g$ band into two non-degenerate bands. One of them is completely empty 
 (the peak just above E$_F$) and the other partially filled (the peak at E$_F$). The
 remaining charge partially occupies the antibonding spin up     
  Fe$_B$ bands which have t$_{2g}$ symmetry . The system thus prefers to assume a
 partially inverted structure with respect to the fully inverse one of magnetite where
 only Fe$_B$ antibonding states are at the Fermi level.  
 Due to the shorter Fe$_B$-Fe$_B$ distance of 2.92 ${\AA}$ 
  the metal-metal bonding produces a  larger bandwidth of Fe$_B$ states at E$_F$ as well as 
  a higher electronic population of Fe$_B$ states in (ii) with respect to (i)  as is clear by
  comparison of  the DOS of the minority
  Fe$_B$ states between case (i) and (ii) (see  Figs. \ref{dosa2b} and  \ref{dosa2bii} upper panel). We may 
  therefore conclude that the degree of inversion in  (ii)
  is larger than that in  (i) and we may then argue that the driving mechanism in the inversion of magnetite
  is due to the direct cation(B)-cation(B) bonding.
  
   Moreover, from the bandstructure (Fig. \ref{dosa2bii}) it is clear that
 the spin down  Fe$_A$ band at E$_F$ (lower panel) is much less dispersive than the
spin up Fe$_B$ band (upper panel)
 resulting in a Mott-Hubbard like behavior of the A site.  
 Therefore, we find here
  a particular kind of half-metallic behaviour of the  system with respect to the conventional
  full spin polarized half-metals. 
  In fact, both spin species are here present 
  at the Fermi level but only one of them  ($\downarrow$ t$_{2g}$ Fe$_B$ states)
 is  significantly conducting  while the other ($\uparrow$ e$_{g}$ Fe$_A$ states) 
 is strongly localized. This kind of system is called  {\it transport} half-metal
 as opposed to a conventional half metal where the spin polarization (P) is 100 $\%$.
 We calculated the value of P for the present case according to the definition\cite{nadgorny} 
$P_n={ N_{\uparrow}v_{\uparrow}^n-N_{\downarrow}v_{\downarrow}^n \over N_{\uparrow}v_{\uparrow}^n+ N_{\downarrow}v_{\downarrow}^n }$ ,
where $N$ is the density of states at the Fermi level for the two spin species
and $v$ the respective Fermi velocity. From the calculated band structure
and DOS we derive a value of P$_1$ equal to  35 $\%$ and of  P$_2$ equal to  81 $\%$.
Since the bulk current is  proportional to $N(E_F)v_F^2$ , the  P$_2$ value implies
that 90.5 $\%$ of the current is carried by the spin $\uparrow$ electrons,
(namely the minority spins of Fe$_B$) and a  {\it transport} half-metal
is clearly realized.
 
However, we should recall that the charge carrier concentration in this case is less than that in magnetite
and therefore the metallicity of this system is poorer than the one of magnetite.

Finally, some considerations on the stability of the proposed structures should be done at
 this stage.
Due to the lack of implementation of atomic forces calculation
for an LDA+U potential in the WIEN2k code 
and aware of the fact that the inversion issue in spinels is
not intrinsically due to on-site Hubbard effects, we have estimated the
instability of structures (i) and (ii) respect to inversion 
by evaluating atomic forces within the  GGA potential. 
The structures (i) and (ii)  considered here are the GGA optimized ones
and have not been relaxed after replacement of Zn and Ga with Fe.
In Table \ref{force} we show the moduli of the forces acting on 
Fe$_A$, Fe$_B$ and the oxygen connecting them, which are the most significant ones.
Note the difference in magnitude between cases (i) and (ii) for Fe$_A$ and oxygen
due to the larger degree of inversion of (ii) respect to (i).
Comparing the force directions (not shown in the Table) we see that
the distance between Fe$_A$ and oxygen has to become shorter,
testifying the tendence of both structures to invert. 
Clearly the structure (ii) is significantly more unstable than the structure (i).

\begin{center}
\begin{table}
\caption{Comparison of forces on Fe and O atoms in structure (i) and (ii) (units mRy/a.u.)}
\vspace{.3cm}
\begin{tabular}{c c c c}
\hline
\hline
\hspace{.5cm}      &  Fe$_A$  & Fe$_B$  &  O      \\
\hline
(i) \hspace{.5cm}  &  20.589  & 6.623    & 36.442   \\
(ii)\hspace{.5cm}  &  40.395  & 9.240    & 60.324   \\
\hline
\hline
\end{tabular}
\label{force}
\end{table}
\end{center}

\section{ Conclusions}
In conclusion, we have performed a systematic {\it ab-initio} study of Fe doped \zn to investigate the effect of Fe substitution
on the electronic structure of the parent spinel compound \zn. For an accurate account of the magnetic and transport properties
we considered various exchange correlation functionals and 
 we observed that the electronic properties of the doped system depend significantly
on the exchange-correlation functional used, especially when Fe is doped on the  B cation site or on both A and B cation sites of the spinel
compound. Whereas in the case of Fe doped in the A site both LDA and GGA give qualitatively the same results (close to an ionic picture),
B-site doping leads to different ground state spin of Fe within LDA and GGA. During the course of this study we also
tried to explore the possibilities of obtaining a strongly polarized half-metallic groundstate which is a desirable
property for spintronic applications. We observed that Fe doping in A-site could give rise to a spin polarized current
due to the presence of highly dispersive $s$ band in one spin channel. In B-site, Mn instead of Fe could lead to a half-metallic
state by setting the Fermi level in a comparatively wider $e_g$ band in the spin up channel. Doping both A and B sites with
Fe leads to a so called 'transport half metal' when there is a direct Fe-Fe bonding in the B-sites.

   Finally, recent work on [ZnGa$_2$O$_4$]$_{1-x}$[Fe$_3$O$_4$]$_x$ for various
large Fe doping concentrations, 
 showed the existence of a ferromagnetic phase up to 200K.
 M{\"o}ssbauer spectra were measured and provided a quadrupolar splitting of 0.52 mm/s.   Assuming a nuclear quadrupole moment for $^{57}$Fe of 0.16 barn, as calculated
 in ref. \cite{QFelapw} within the  LAPW method, 
 the resulting electric field gradient (EFG) is 3.12 V/cm$^2$. 
 Alternatively, assuming a nuclear quadrupole moment of 0.2 barn, 
 as calculated in ref. \cite{QFehf}, the EFG is 2.5 V/cm$^2$.
 When Fe is substituted into an A site no EFG develops
 due to the spherical local symmetry at the nucleus.
 Instead, in the case of Fe doped into B site the optimized structure provided us with
 an EFG at the Fe site of 2.11 V/cm$^2$, which 
 compares fairly well with the experimental values.
 In the doping case of Fe into A and B sites we didn't perform
 a structure optimization, and,  since EFG is quite sensitive to
 structural changes, we obtain quite  different values with
 respect to the experimental ones.
 
   The M{\"o}ssbauer experiment of Risbud {\em et al.} detected the presence
 of only Fe$^{3+}$.  This result would discard the option of  Fe doped in the A site
 in favor of Fe doped on the B site since from our  calculations  Fe   in the A site
 takes up  an oxidation state close to 2+ and not to 3+.
 In fact, in  ZnGa$_2$O$_4$  the site preference of Zn$^{2+}$ towards 
the tetrahedral sites is much stronger than the Fe one \cite{navrotsky} 
and this can explain why Risbud {\em et al.} detected the presence 
of only Fe$^{3+}$, which probably occupies predominantly B sites. 
Alternatively, Mg$^{2+}$ has a weaker preference for the A site  than Ga and Fe,
therefore replacing Zn$^{2+}$ with  Mg$^{2+}$  can provide a practically easier way
to populate A sites with Fe,  and MgGa$_2$O$_4$ would be a better host candidate for the
properties presented here concerning Fe doping on the A site.
\vspace{0.5cm}

\noindent {\bf Acknowledgements}
We would like to thank R. Seshadri for useful discussions and the German Science
 Foundation for financial support. We would also like to thank P. Blaha for helpful 
 discussion regarding the code WIEN2k.



\end{document}